\documentclass{ecai} 

\usepackage[T1]{fontenc}
\usepackage[utf8]{inputenc}
\usepackage{listings}
\usepackage{color}
\usepackage{amsmath}
\usepackage{hyperref}
\usepackage{enumitem}
\usepackage{graphicx}
\usepackage{comment}
\usepackage{caption}
\usepackage{multirow}
\usepackage{float}
\usepackage{comment}
\usepackage[english]{nomencl}
\usepackage{xcolor}
\usepackage{datetime2}
\usepackage{tabularx}

\usepackage{latexsym}
\usepackage{amssymb}
\usepackage{amsmath}
\usepackage{amsthm}
\usepackage{booktabs}
\usepackage{enumitem}
\usepackage{graphicx}
\usepackage{color}

\newcommand{\ignore}[1]{}
 \newcounter{todocounter}
\newcommand{\ma}[1]{\stepcounter{todocounter}
  {\color{red} Mehwish: \thetodocounter: #1}}

\newcommand{\zg}[2]{
  {\color{blue} Zeinab: }}

\newcommand{\BibTeX}{B\kern-.05em{\sc i\kern-.025em b}\kern-.08em\TeX}

\begin{document}


\begin{frontmatter}


\paperid{123} 


\title{Enriching Taxonomies using Large Language Models}


\author{\fnms{Zeinab}~\snm{Ghamlouch}\thanks{Corresponding Author. Email: zeinab.b.ghamlouch@gmail.com}}
\author{\fnms{Mehwish}~\snm{Alam}\orcid{0000-0002-7867-6612}\thanks{Email: mehwish.alam@telecom-paris.fr}} 


\address[A]{Télécom Paris, Institut Polytechnique de Paris, France}

\begin{abstract}
Taxonomies play a vital role in structuring and categorizing information across domains. However, many existing taxonomies suffer from limited coverage and outdated or ambiguous nodes, reducing their effectiveness in knowledge retrieval. To address this, we present Taxoria, a novel taxonomy enrichment pipeline that leverages Large Language Models (LLMs) to enhance a given taxonomy. Unlike approaches that extract internal LLM taxonomies, Taxoria uses an existing taxonomy as a seed and prompts an LLM to propose candidate nodes for enrichment. These candidates are then validated to mitigate hallucinations and ensure semantic relevance before integration. The final output includes an enriched taxonomy with provenance tracking and visualization of the final merged taxonomy for analysis. 

\end{abstract}

\end{frontmatter}


\section{Introduction}

Taxonomies play a crucial role in structuring and hierarchically categorizing information for better organization and clarity. They are widely used across various domains including web data, e-commerce, digital libraries, general purpose knowledge graphs (such as YAGO~\cite{SuchanekABCPS24}, DBpedia, Wikidata), and biomedical domain (such as MeSH, SNOMED-CT, etc.) to support efficient information retrieval, semantic reasoning, and knowledge management. However, many existing taxonomies, such as Schema.org, exhibit limitations in their coverage and granularity. These include missing, ambiguous or outdated nodes. Such gaps hinder the effectiveness of taxonomy-driven systems and reduce their ability to reflect the evolving nature of knowledge. Enriching a taxonomy entails enhancing its structural depth and contextual relevance by introducing new, semantically meaningful categories without disrupting its original hierarchy.

Many existing taxonomies are either derived from textual data or manually built in response to user needs. Large Language Models (LLMS) on the other hand, are trained on large amount of data on the web, storing plethora of knowledge. GPTKB~\cite{corr/abs-2411-04920} is one of the knowledge graphs which given a seed entity extracts the knowledge of an LLM about that entity in the form of triples. It also extracts the taxonomy of the LLMs which the authors claim to be close to the taxonomy of Schema.org. In contrast, while GPTKB explicitly extracts the internal taxonomy of an LLM, our approach uses a seed taxonomy to guide the LLM's output.


In this work, we propose a novel taxonomy enrichment pipeline, Taxoria, which enables live enrichment of a given taxonomy using the inherent knowledge of an LLM. It leverages the existing taxonomy as a seed for generating candidate nodes for enrichment using zero-shot prompting. These suggestions are post-processed and validated using various techniques to mitigate hallucinations and over-generations. After validation, the node is merged into the original taxonomy with the provenance information. Additionally, it also provides visualization of the merged taxonomy allowing deeper analysis. The live demo is available online\footnote{\url{https://taxoria.r2.enst.fr/}}, the source code is available through Github\footnote{\url{https://github.com/zeinabGhamlouch/Taxoria}}, along with the video demonstration\footnote{\url{https://www.youtube.com/watch?v=7-9Hrg-awGo}}.

This paper is structured as follows: Section~\ref{sec:related_work} places the tool in reference to existing studies. Section~\ref{sec:architecture} gives the overall architecture of Taxoria while Section~\ref{sec:use-case}  discusses the use-case of real-world taxonomies. Finally, Section~\ref{sec:discussion} highlights the limitations and discusses the future directions.

\section{Related Work}
\label{sec:related_work}
Traditional approaches to taxonomy enrichment, such as manual curation, rule-based systems, and statistical models face critical challenges. YAGO4.5~\cite{SuchanekABCPS24} performs manual taxonomy refinement with expert intervention using Schema.org as the base taxonomy, however, manual enrichment is resource-intensive. Rule-based methods lack adaptability and generalizations~\cite{Angermann22}, and statistical techniques often fail to capture deeper semantic relationships. On the other hand, LLMs capture large body of knowledge including taxonomic knowledge. Various studies have utilized LLMs for taxonomy expansion~\cite{10.1145/3709007} and taxonomy refinement~\cite{conf/cikm/PengBA24}. In~\cite{conf/cikm/PengBA24}, the authors use a combination of graph mining techniques and LLMs, where LLM is used for verifying the links between the nodes. Taxoria is close to the task of taxonomy expansion, however, it uses the internal knowledge of an LLM for the expansion.

\section{Overall Architecture}
\label{sec:architecture}
\begin{figure}
    \centering
    \includegraphics[width=\linewidth]{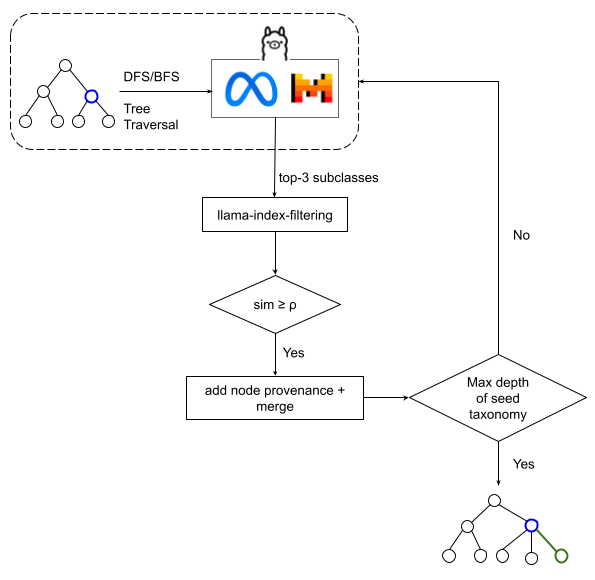}
     
    \caption{Overall architecture of Taxoria. The blue node represents the current node in the seed taxonomy for which the direct children are generated and the green node represents the newly added node after the merging process.}
   \label{fig:taxoria}

   \vspace{0.75cm}
\end{figure}

Figure~\ref{fig:taxoria}, shows the overall architecture of Taxoria. This automated pipeline focuses on enriching the taxonomy based on the internal taxonomy of an LLM. The enrichment process involves the following key steps: starting with a seed taxonomy, the system prompts the LLM for extracting the internal taxonomy of different open-source LLMs retaining only relevant information.

\paragraph{Taxonomy Traversal Strategy.} 
 
We employ two primary taxonomy traversal strategies:

\begin{itemize}
    \item \textbf{Breadth-First Search (BFS).}  All nodes at a given level are expanded before proceeding to the next level. 

    \item \textbf{Depth-First Search (DFS).} A complete branch is expanded before moving to adjacent branches.  
\end{itemize}

\paragraph{Prompting LLM.} 

Zero-shot prompting is used for generating the classes from an LLM using seed node from a taxonomy traversed using one of the previously defined strategies. For a given node only the top-3 direct children were generated which use the internal scoring of an LLM with a constraint to generate an output in a JSON format. 

\paragraph{Mitigating Over-generations LLM.} 

Despite their capabilities, LLMs may produce suggestions that are irrelevant, redundant, or inconsistent due to their tendencies to hallucinate. To address this, we apply the following strategies:

\begin{itemize}
    \item \textbf{Defining Stopping Criteria.} To prevent over-generation and avoid deeper taxonomy, we define stopping criteria based on the maximum depth of the original taxonomy. Newly generated nodes are restricted to extend the hierarchy by at most one additional level beyond the existing maximum depth.
    
    \item \textbf{Filtering based on Semantic Similarity.} We compute the semantic similarity between a generated node and its parent using Word2Vec embeddings. A node is merged into the taxonomy if its cosine similarity to the parent node is greater than a certain threshold $\rho$ (here $\rho \geq 0.9$ within the range [0,1]). For computing the semantic similarity Word2Vec and LlamaIndex were used for ensuring n-gram level as well as contextual level similarity respectively. 
    


\end{itemize}








\subsection{Merging process}

\begin{figure*}
    \centering
    \includegraphics[width=\linewidth]{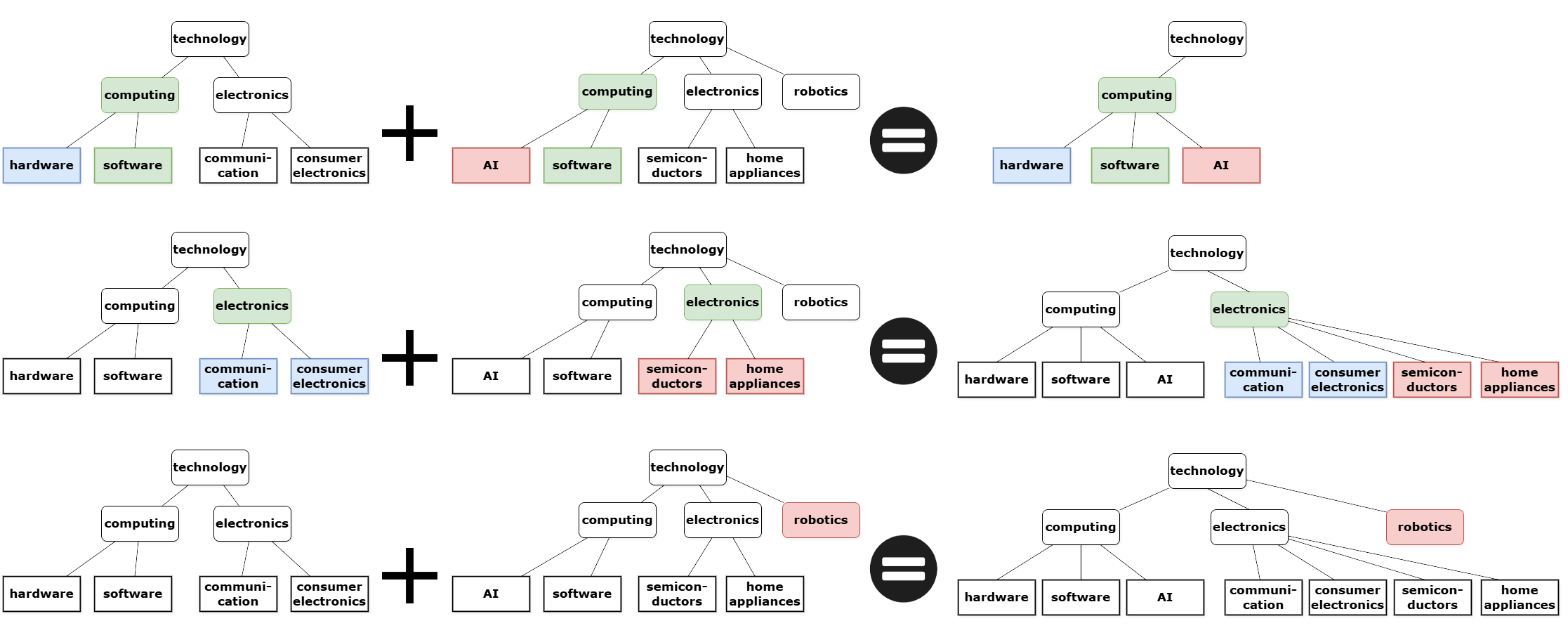}
    \caption{This figure illustrates an example of merging two hierarchical taxonomies in the technology domain. Nodes common to both input taxonomies are shown in \textbf{green}, while nodes unique to each taxonomy are marked in \textbf{blue} and \textbf{red}, respectively.}
    \label{fig:merging}
    \vspace{0.5cm}
    \end{figure*}

Figure~\ref{fig:merging} shows the overall merging process. It integrates new nodes generated by the LLM into the existing taxonomy while avoiding redundancy and preserving the hierarchical structure. The process consists of the following steps:

\paragraph{Node Generation.}
New candidate nodes are generated by the LLM based on the current state of the taxonomy. These nodes are proposed as direct children of existing nodes or as deeper subtrees, depending on the tree traversal strategy used.
    
    \paragraph{Validation of Node Relevance.}
Each generated node is evaluated using the filtering mechanisms described previously to ensure conceptual relevance, avoid redundancy, and maintain structural coherence.
   
    \paragraph{Merging Nodes.}
    The system handles node integration based on whether the generated node already exists in the taxonomy:
    \begin{itemize}
        \item If the node already exists with the same name, the system recursively compares and merges their children. The original children are preserved, and any new children are appended, avoiding duplication.
        \item If no matching node is found, the new node is directly added as a child to the appropriate parent node in the taxonomy.
    \end{itemize}

\paragraph{Adding Class Provenance.} 
To distinguish between original and newly generated nodes, we introduce a \textbf{source key} attribute for tracking the provenance of each node:
\begin{itemize}
    \item \textbf{Original Nodes.} All nodes that were part of the initial taxonomy retain a source key with the value ``original-taxonomy", indicating that they originate from the original taxonomy.
    \item \textbf{Newly Merged Nodes.} All nodes generated and added through the enrichment process are assigned a source key with the value ``llm-generated", signifying that they were derived from the LLM-based pipeline.
\end{itemize}

This explicit tracking mechanism ensures transparency and traceability of enriched content.

\subsection{Implementation Details} 

Taxoria implements open source LLMs, i.e., LLAMA 3, LLAMA 3.2, and Mistral. Each of these LLMs are chosen based on their characteristics, i.e.,

\begin{itemize}
    \item 
\textbf{LLaMA 3 \& 3.2} are chosen for their strong language generation performance and low-latency while generating responses. 
\item \textbf{Mistral} is selected for its compact model size and efficient inference speed. 
\end{itemize}

Ollama\footnote{\url{https://ollama.com/}} is chosen to run the LLMs on the backend. This choice is motivated by the fact that, the smaller LLMs can be run locally offering control over the environment and ensuring data privacy which is one of the most important concerns in various industry/domain oriented applications. Additionally, it also provides access to a number of LLMs and not only the ones chosen in this work allowing flexibility while building applications. Taxoria uses NVIDIA GTX 1080 GPU, LLMs with higher number of parameters can be used upon the availability of higher computational resources.  The use of Ollama also allows to run Taxoria on local machine.


\section{Use-case and Impact}
\label{sec:use-case}

This section discusses the application of Taxoria on real-taxonomies leading to a discussion on the broader impact of the tool.

\paragraph{Real-World Taxonomy Enrichment.}

We tested Taxoria in the real setting with real world taxonomies. These taxonomies span diverse domains, including general knowledge (Schema.org\footnote{\url{https://schema.org/}}), e-commerce (eBay), and geographical information (GeoNames)~\cite{SunHSXYDTC24}. Schema.org is tailored for structured metadata on the web, eBay for hierarchical product categorization, and GeoNames for geographic entity classification.  Table~\ref{tab:taxonomy-stats} shows the statistics of the taxonomies and the number of newly LLM-generated classes.

\paragraph{Analysis.} While enriching the previously described taxonomies, we encountered predictions which were irrelevant or were entities and not the classes. However, these strategies are not directly integrated into the demo in order to ensure it functions within limited computational resources.

\begin{itemize}
    \item \textbf{LLM-Based Relevance Filtering.} In this scenario, we use LLM-as-a-judge for evaluating the relevance of a generated node with respect to its parent. The node is retained only if its relevance score meets or exceeds the average similarity observed in the original taxonomy.
    
    \item \textbf{Validation against an External Knowledge Graph.} In some cases, LLMs generate instances (entities) of a class, e.g., \texttt{Paris} which is an instance of a class \texttt{City}. During this validation process, we use external knowledge graph such as Wikidata~\cite{VrandecicK14} to verify if the generated node is an instance or a class. For doing so, Wikidata is queried to check the existence of ``instance of" (P31) property. If the property is not associated to generated node, the node is retained otherwise it is discarded.
\end{itemize}

\begin{table}[h!]
\centering
\begin{tabular}{@{}lcccc@{}}
\hline
\textbf{Taxonomy} & \textbf{\# Classes} & \textbf{Max Depth} & \textbf{\# New Classes} & \textbf{New Max Depth} \\
\hline
Schema.org    & 1143 & 6  & 4216 & 7\\
eBay          & 596   & 4  & 1106 & 5\\
GeoNames      & 690   & 3  & 1138 & 4\\
\hline
\end{tabular}
\caption{Statistics of the Taxonomy. The \# new classes are reported only for Llama 3.2.}
\label{tab:taxonomy-stats}
\end{table}



\paragraph{Impact.}

Taxonomies defined in general purpose knowledge graphs such as DBpedia, Wikidata, and YAGO have various limitations. The taxonomy of DBpedia is semi-automatically generated which is incomplete. Wikidata's taxonomy is often disorganized as a result of its crowd-sourced contributions. However, the taxonomy of YAGO has evolved over time. It started with the use of WordNet hierarchy, which then evolved to using Wikidata taxonomy. The taxonomy of the current version of, YAGO 4.5, is manually curated based on Schema.org. This tool can further be used for creating richer taxonomies for these knowledge graphs. The live enrichment of this taxonomy can also serve as an agent for these knowledge graphs, enabling dynamic integration of new information.

Taxonomies related to a certain domain such as biomedicine are vast and complex, often maintained separately. Merging tools can help integrate taxonomies across sources, while LLMs can be leveraged to enrich them with up-to-date biomedical knowledge, especially where scientific field is evolving fast and it is not possible to manually update the taxonomy.

\section{Discussion and Future Directions}
\label{sec:discussion}

While applying Taxoria on the real-world taxonomies, we observed that at a certain level of the taxonomy instead of generating classes as the subclasses of the node from the seed taxonomy, the LLM starts generating instances of those classes. In such a scenario we decided to use external knowledge graphs such as Wikidata for grounding the LLM generation by using the ``instance of" relation using SPARQL queries. However, this led to various issues, i.e., given that Wikidata is a semi-automatically generated and is a crowd sourced effort, it contains many inconsistencies and erroneous type information. As a future direction, we would extend the tool to simultaneously perform entity type prediction~\cite{BiswasPPSA22} for a certain domain related to seed taxonomy. 

The current visualization of Taxoria does not support edits. In the future, this could be extended to allow human evaluation within the tool by allowing live user edits. Further extensions would support tasks such as taxonomy induction, expansion, and refinement. Additionally, robust validation strategies will be explored to ensure the accuracy and reliability of the merged taxonomies, including both automated evaluation metrics and human-in-the-loop approaches.

\section*{Acknowledgement.}

This publication is based upon work from COST Action CA23147 GOBLIN - Global Network on Large-Scale, Cross-domain and Multilingual Open Knowledge Graphs, supported by COST (European Cooperation in Science and Technology, \url{https://www.cost.eu}).







\bibliography{mybibfile}

\end{document}